# Investigating Decentralized Management of Wearable Device Data for Health

Tao Zhou
Department of Information, Artificial Intelligence and Data Science
Daiichi Institute of Technology
Tokyo, Japan
21te477@ueno.daiichi-koudai.ac.jp

Tianle Hu
Department of Information, Artificial Intelligence and Data Science
Daiichi Institute of Technology
Tokyo, Japan
24te513@ueno.daiichi-koudai.ac.jp

Hong Chen
Department of Information, Artificial Intelligence and Data Science
Daiichi Institute of Technology
Tokyo, Japan
h.chen@ueno.daiichi-koudai.ac.jp

***Abstract*—Wearable devices have wide applications in healthcare, and an increasing amount of health-related information is being collected and stored in the cloud. However, the existence of data silos hinders users and professionals from obtaining a comprehensive view of health. Combining data sources from different wearable devices with other personal data in a secure manner has great potential benefits for individuals, healthcare professionals, and researchers. Decentralized architectures such as UPOD offer a solution that enables users to better control the data in their UPODs. This study proposes the Ubi-Health framework, integrating sensor data from various types of wearable devices into UPOD, and creates Pod App, a proof-of-concept decentralized mobile health application. Through the Pod App, individuals can make more informed decisions about their health and well-being. Researchers can utilize data from wearable devices to analyze users' health trends and identify potential risk factors.**

## I. INTRODUCTION

In the current healthcare sector, wearable devices such as smartwatches and fitness bands are becoming increasingly popular. These devices can monitor health-related metrics in real-time, such as heart rate, blood pressure, and blood oxygen levels, providing valuable health insights to users and healthcare professionals. With the advancement of sensor and operating system hardware technologies, wearable devices are gradually expanding their functions to include more diverse forms and more accurate physiological metrics. These sensors are moving towards higher precision, continuity, and comfort, making significant contributions to improving personalized healthcare[1].

As shown in Figure 1, due to the use of different communication protocols (such as TCP/IP and Bluetooth) and proprietary data formats between different manufacturers, the data generated by these devices is often incompatible, creating data silos. Users' health data is scattered across different platforms, making it difficult to integrate and share across devices. This fragmentation not only affects users' comprehensive understanding of their own health status but also limits healthcare professionals' access to and analysis of long-term patient health data.

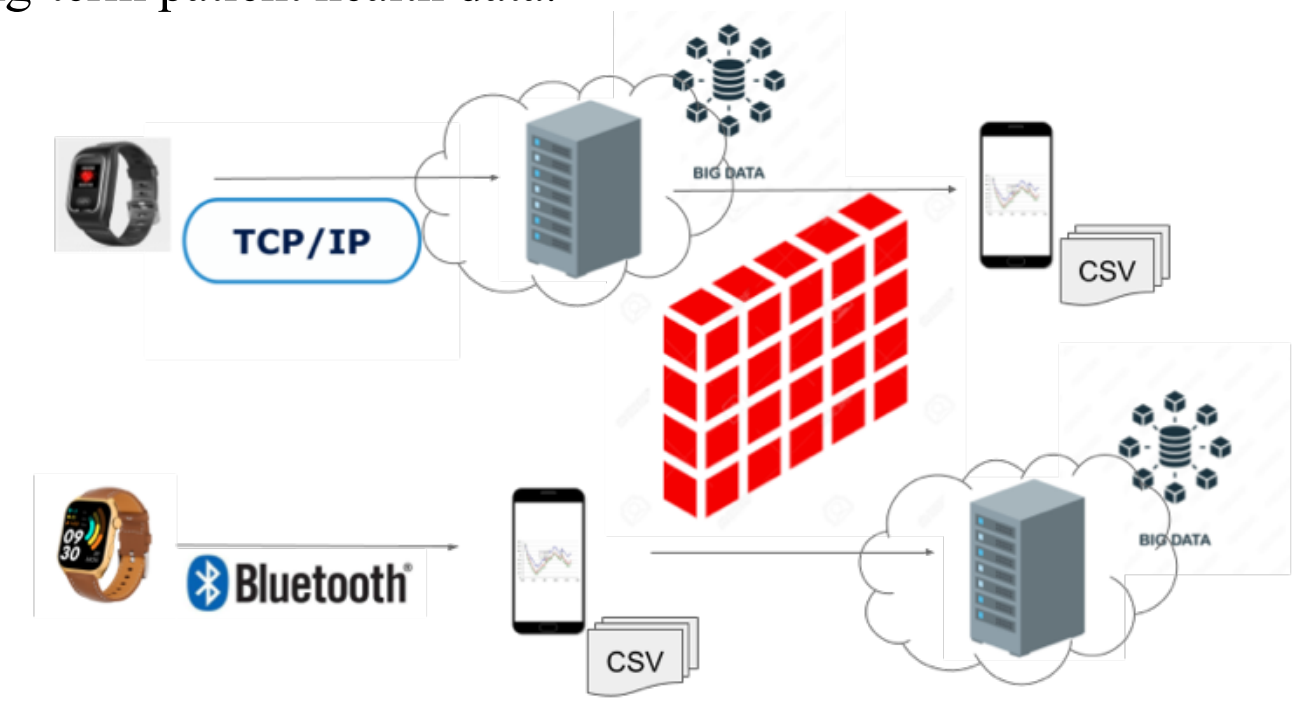

Figure 1. Personal data from wearable devices exhibits the phenomenon of data silos.

## II. RELATED WORK

### A. Smart wearables with various sensors

In recent years, with the widespread application of wearable devices in health monitoring, more and more research has focused on the role of these devices in real-time health data monitoring and personalized healthcare. [3] mentions the nature of physiological signals, desired life parameters, the role of smart wearables, the selection of wearable devices, and design considerations of wearable devices for early detection of health conditions. It also provides critical data support for healthcare professionals, helping to detect health issues early and formulate personalized treatment plans. The emergence of these devices has significantly transformed the traditional healthcare model, driving the rapid development of remote health monitoring and personalized medicine. However, with the widespread use of these devices, issues such as data accuracy, device reliability, and data privacy protection have become central topics of research. Shifts the control of electronic healthcare enterprises from the organizational level to the patient level [4]. This aims to give patients more control and authority over the decisions regarding their protected health information/electronic health records. A significant amount of work and effort is needed to better assess the feasibility of this major shift in electronic healthcare enterprises. Existing research can be naturally categorized based on the technologies used. These include data anonymization/pseudonymization and access control mechanisms, primarily aimed at preserving data privacy.

At the same time, with the increasing amount of data collected by wearable devices, effective data management and sharing have become key research focuses. Ensuring that the data generated by these devices is effectively protected during secure sharing and transmission is crucial. Additionally, ensuring secure cross-platform data sharing and management is also a current research focus. During data transmission and storage, multi-layer encryption and privacy protection protocols must be implemented to reduce the risk of data breaches.

### *B. Disease prediction using machine learning methods*

Machine learning plays a crucial role in processing large amounts of data generated by wearable devices and in disease prediction. For conditions such as cardiovascular diseases and diabetes, machine learning algorithms can analyze health data from wearable devices to help achieve early detection and personalized treatment strategies. Optimizing algorithms is essential for improving prediction accuracy, and research has shown how various advanced machine-learning methods enhance the performance of health data prediction models, making them more suitable for clinical applications. Moreover, the scheme allows a user to share his data with other users without requiring the security constraints implemented by the cloud service providers[5]. At the same time, [6] mentions shifting control of electronic healthcare enterprises from the organizational level to the patient level. This aims to give patients more control and authority over the decisions regarding their protected health information/electronic health records.

### *C. Mobile applications for healthcare*

Mobile applications make health data generated by wearable devices more accessible to patients and healthcare providers. These applications allow users to track their health and provide feedback based on real-time data from wearable devices. Additionally, these apps facilitate remote consultations and disease management, contributing to improved patient health outcomes. Currently, many studies focus on the development and optimization of health-related mobile apps, emphasizing their importance in enhancing user experience and expanding access to healthcare services. For example, [7] mentions that blockchain, characterized by transparency, immutability, trustlessness, and decentralization, can help establish a secure healthcare data exchange network. In the context of [8], they present a 2 × 2 Partnership-Identity Exposure Matrix, illustrating implications in four different scenarios and providing recommendations for improving privacy protection.

### *D. Personal online data store*

As wearable technology continues to advance, the secure sharing and management of data have become a key research focus. Recent developments have proposed the use of decentralized frameworks (such as Solid), allowing users to store their personal data in their own Pod instead of relying on centralized servers. This approach enhances privacy protection by giving individuals greater control over their health data, reducing the risk of data breaches, and ensuring secure data sharing across different platforms [9]. The combination of the Solid framework and ubiquitous personal online data store (UPOD) system aims to reduce reliance on centralized servers, lowering potential privacy risks while improving the interoperability and security of cross-platform data transmission [10]. Additionally, our research demonstrates that the application of standardized data protocols and distributed data management strategies further enhances data interoperability and privacy protection. By achieving a unified data transmission format across different systems, users' health data can be securely and effectively shared and utilized in cross-platform environments[11, 12]. This approach not only simplifies data management processes but also strengthens users' trust in health data management systems[13].

## III. PROPOSED SYSTEM

This section provides a detailed explanation of the decentralized wearable device system framework designed in this study. In this framework, two main types of data transmission are used to handle health data generated by smartwatches, and personal data management and analysis are implemented based on UPOD.

### *A. Ubi-Health Framework*

This study proposes a decentralized Ubi-Health framework. The framework is based on the ActivityPub protocol, integrating data from various wearable devices such as smartwatches and fitness bands, as well as social networks. This data is stored in the ubiquitous personal online data store (UPOD), ensuring complete separation of data and applications, and assigning different roles based on data types.

Wearable devices, such as smartwatches, fitness bands, and smart glasses, come in various forms, with their main communication methods divided into TCP/IP and Bluetooth. Based on different communication methods, wearable devices can be roughly divided into two categories: devices that communicate via TCP/IP and devices that communicate via Bluetooth. The data transmission paths mainly include two methods, as shown in Figure 2. The first path involves transmitting data from wearable devices to a dedicated server, after which the data is stored in UPOD. The second path transmits the data directly from the wearable device's application to UPOD. UPOD, as a decentralized platform for personal data storage and management, ensures that users have full control over their data.

Pod App serves as the interface for users to interact with UPOD, allowing them to view the data stored in UPOD. Through Pod App, users can access their health data in real time and can analyze and track their historical health records at any time. We hope that more health applications and systems for wearable devices can be built on UPOD, as many use cases can be developed based on the work we have done on UPOD. For example, an application similar to Pod App can be created that uses federated learning to train and test models using health data from multiple users.

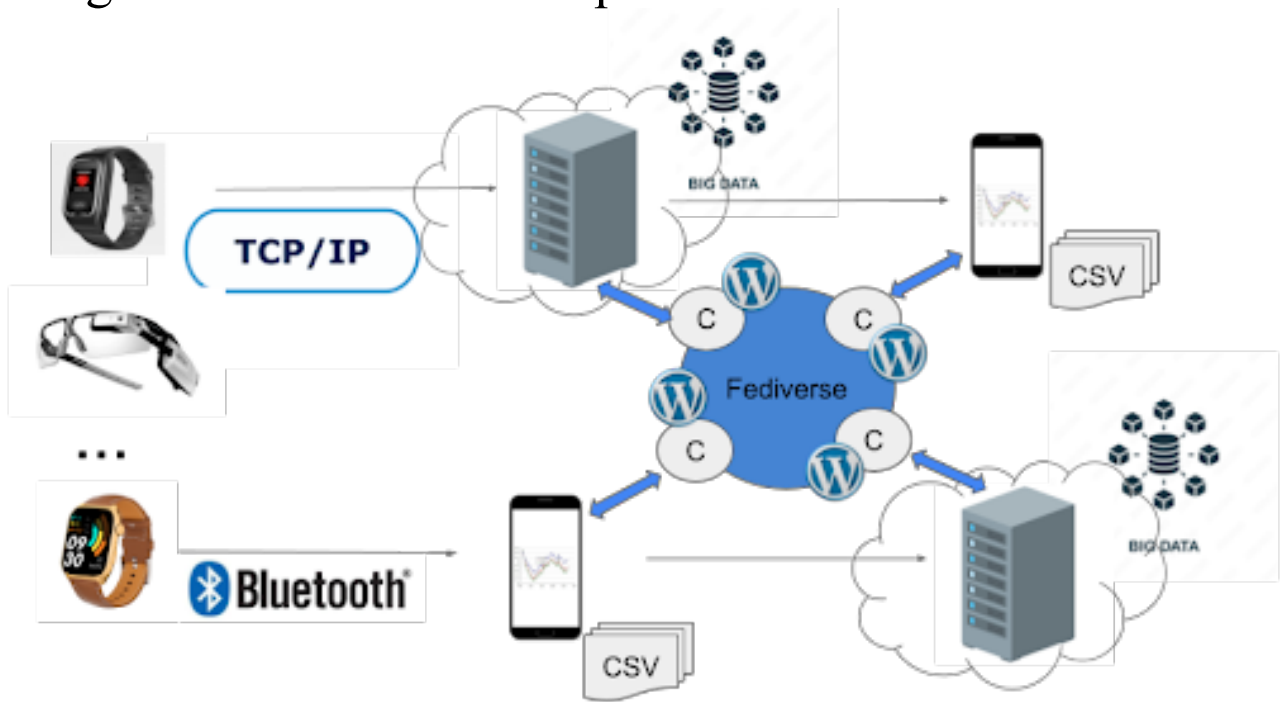

Figure 2. Proposed Framework

## B. *Integration with the Fediverse*

The Fediverse is a global social network composed of interconnected servers, allowing people to communicate across different platforms. UPOD does not own the Fediverse. UPOD is just one of many servers that have joined the Fediverse. Pod App is an application of UPOD, and it is compatible with the Fediverse through integration with the open social network protocol ActivityPub. Pod App will become one of the Fediverse servers, enabling content to flow between Pod App and other servers.

## C. *Data Sharing*

ActivityPub is an open-source, decentralized social networking protocol currently used by applications such as Mastodon, Pixelfed, and the streaming service PeerTube. It provides a client-to-server API for creating, updating, and deleting content, as well as a federated server-to-server API for delivering notifications and subscribing to content.

As shown in Figure 3, the UPOD application supports bidirectional following for users to access relevant data, addressing data sharing, privacy, and security concerns. By integrating AI big data algorithms, the framework can perform health analysis based on physiological and activity data. Using big data disease models and AI algorithms, users can input their existing physical data, and by continuously wearing the device, the learning model provides professional recommendations to improve health conditions.

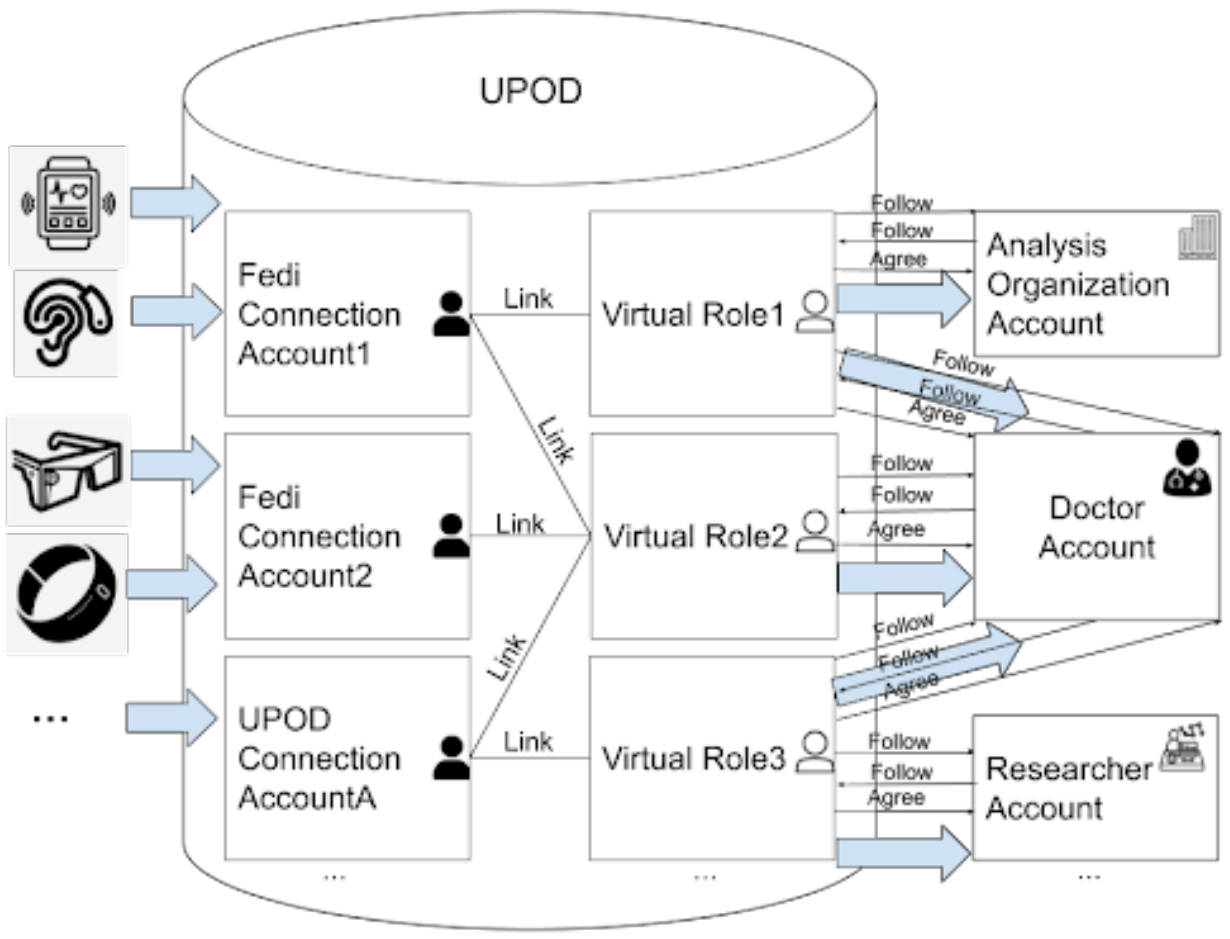

Figure 3. Data Sharing

## D. *Machine Learning Models*

While storing data, UPOD also collaborates with other servers to conduct joint analysis using federated learning technology. During this process, different users' health data can be used to train and optimize machine learning models, but due to the nature of federated learning, the data is not leaked or stored in a centralized manner outside of UPOD, ensuring data privacy protection. Federated learning allows different UPOD instances to collaboratively train models without exposing users' raw data to any third-party servers. Ultimately, the analysis results completed through collaboration between UPOD and other servers are fed back to the Pod App via UPOD. Users can view these analysis results via the Pod App and make appropriate health management decisions based on the feedback from their health data. For example, users can obtain heart rate analysis reports, sleep quality assessments, or receive exercise recommendations through the Pod App.

# IV. EXPERIMENT

We further built a prototype to evaluate the proposed framework, as shown in Figure 4. In the prototype of this experiment, we focused on smartwatches, testing the integration of data sources from various wearable devices with other personal data in a secure manner, demonstrating significant potential benefits for individuals, healthcare professionals, and researchers.

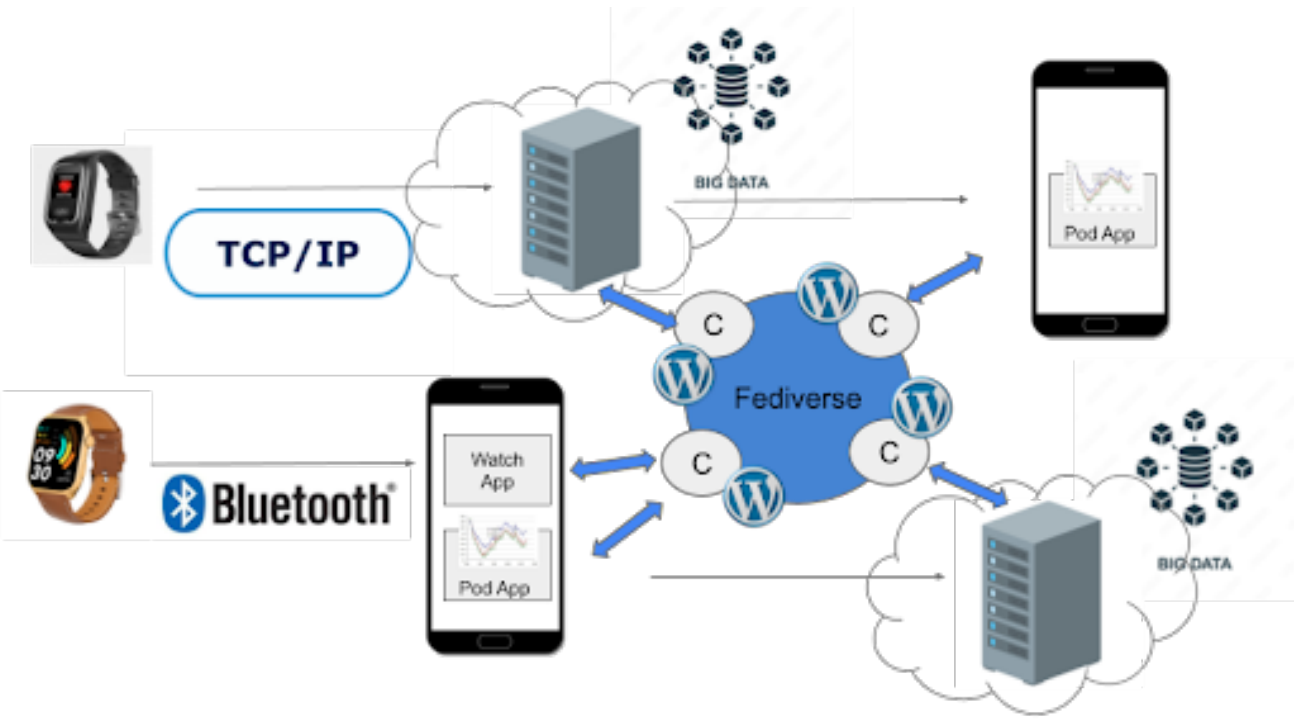

Figure 4. Prototype demonstration of the Ubi-Health framework

### A. Data Collection

Under the TCP/IP and Bluetooth protocols, we send commands from the server to the smartwatch to request data from the wearable device. The data is then sent to the server and shared with UPOD.

In the case of the first path, we configured a server for devices that use the TCP/IP protocol. The Pod App sends commands to the smartwatch through the server to request data from the wearable device, and the data is stored in UPOD via the UPOD API.

As for the second path, we developed a Watch App to collect data from Bluetooth-enabled wearable devices. The Pod App sends a data request to the smartwatch, and the smartwatch sends the data to the Watch App. The Watch App stores the data in UPOD via the API.

We implemented specific functions by developing different modules within the Watch App. We developed corresponding modules such as the blood pressure module, blood oxygen module, and activity module, to handle data in their respective formats. First, we request the required data. The health module sends a request to the Bluetooth module, which forwards the request to the smartwatch. The watch receives the command and sends the data back in the corresponding format to the Bluetooth module, which then forwards the data to the respective module. In each module, we wrote code to connect and transmit data to UPOD via the API.

### B. Health Management Backend and Doctor Diagnosis

After UPOD receives the big data, it shares the data with doctors through authorization, allowing doctors to perform health assessments based on their professional expertise. Through dynamic monitoring, the health status of the body, whether healthy, sub-healthy, or diseased, can be understood, providing a theoretical basis for further intervention and treatment. However, the reliability of the devices and the quality of the data are critical factors that affect the accuracy of health assessments.

Through dynamic monitoring, we can understand the health status of the body, whether it is healthy, sub-healthy, or diseased, and provide a theoretical basis for further intervention and treatment.

The reliability of devices varies greatly, partly due to improper use by the users. When users present their health data and expect health experts to make diagnoses based on the data they collected, this becomes a major concern for health experts. Currently, our system is unable to detect the quality of sensor readings. However, it does tag the source of the data. Knowing the source of the collected data helps health experts assess the quality of the data.

### C. Feedback via Pod App

The proposed framework has been validated in health management and applications by collecting physiological information through smartwatches. Below is a brief introduction to some specific applications.

Each smartwatch is connected to an individual's UPOD, and UPOD sends the collected data and health assessments to the individual via the Pod App, forming a health record and generating a daily health report. If the data is abnormal and reaches a warning level, the individual will be notified via a smartphone pop-up or SMS.

Currently, our application records two heart rate readings and marks the source of each reading, providing guidance to users in case of discrepancies. In future versions, our application will offer users the option to choose which sensor should be used for real-time analysis and feedback. As data sources increase, it will be necessary to differentiate sources based on their reliability, quality, and level of trust.

## V. Conclusion

Data ownership and privacy are particularly important, especially when it comes to personal wearable device data. This study provides a general decentralized application framework capable of offering these functionalities. In this paper, we explore the potential of UPOD as a decentralized management tool for wearable device data. We developed functionalities to integrate wearable device sensor data into UPOD and created Pod App, a proof-of-concept decentralized mobile health application. We evaluated these contributions through the constructed prototype. By introducing wearable device data into the UPOD ecosystem, we hope this exploration will continue to progress.

Our vision is that by integrating UPOD, Pod App, and wearable devices, users will be able to securely and privacy-friendly manage and share their wearable device data. Pod App not only allows users to interact with the global social network but also, through the ActivityPub protocol, enables sharing of health data from wearable devices such as smartwatches with other compatible servers and trusted services. We aim to provide a unified experience across platforms and devices, breaking data silos, and empowering users with greater control and autonomy over their health data while enjoying the benefits of a distributed social network.

Future work will focus on expanding the capabilities of the Ubi-Health framework by integrating more advanced AI algorithms and exploring its applications in broader healthcare environments. Additionally, further research is needed to enhance the scalability of the framework and address challenges that may arise during the adoption of decentralized health data management systems. Overall, the Ubi-Health framework represents a significant advancement in the evolution of wearable health technology, offering a more integrated, secure, and user-centered approach to personal health management.

## References

[1] Smith, Aaron Asael, Rui Li, and Zion Tsz Ho Tse. "Reshaping healthcare with wearable biosensors." *Scientific Reports* 13.1 (2023): 4998.

[2] Y. Ni, T. Zhou, X. Gao, and H. Chen, "Design and Implementation of a UPOD for Decentralized IoT Data Platform Based on the ActivityPub Protocol," *In 2023 IEEE Intl Conf on Cyber Science and Technology Congress(CyberSciTech)*, 2023, pp. 0825-0830.

[3] Nahavandi, Darius, et al. "Application of artificial intelligence in wearable devices: Opportunities and challenges." *Computer Methods and Programs in Biomedicine* 213 (2022): 106541.

[4] Li, He, et al. "Examining individuals' adoption of healthcare wearable devices: An empirical study from privacy calculus perspective." International journal of medical informatics 88 (2016): 8-17.

[5] M. Tanveer, A. U. Khan, M. Ahmad, T. N. Nguyen and A. A. A. El-Latif, "Resource-Efficient Authenticated Data Sharing Mechanism for Smart Wearable Systems," in IEEE Transactions on Network Science and Engineering, vol. 10, no. 5, pp. 2525-2536, 1 Sept.-Oct. 2023, doi: 10.1109/TNSE.2022.3203927.

[6] M. A. Sahi et al., "Privacy Preservation in e-Healthcare Environments: State of the Art and Future Directions," in IEEE Access, vol. 6, pp. 464-478, 2018, doi: 10.1109/ACCESS.2017.2767561.

[7] H. Jin, Y. Luo, P. Li and J. Mathew, "A Review of Secure and Privacy-Preserving Medical Data Sharing," in IEEE Access, vol. 7, pp. 61656-61669, 2019, doi: 10.1109/ACCESS.2019.2916503.

[8] Banerjee, Syagnik, Thomas Hemphill, and Phil Longstreet. "Wearable devices and healthcare: Data sharing and privacy." The Information Society 34.1 (2018): 49-57.

[9] H. Chen, "Ubi-Care: a Decentralized Ubiquitous Sensing Healthcare System for the Elderly Living Support," In 2019 IEEE Intl Conf on Cyber Science and Technology Congress (CyberSciTech), 2019, pp. 543-547.

[10] T. Zhou, Y. Ni, X. Gao, and H. Chen, "Smart Home Integration and Data Sharing with ActivityPub and UPOD," In 2023 IEEE Intl Conf on Cyber Science and Technology Congress(CyberSciTech), 2023, pp. 0056-0062.

[11] Q. Jin, B. Wu, S. Nishimura, and A. Ogihara. "Ubi-Liven: a human-centric safe and secure framework of ubiquitous living environments for the elderly." In 2016 International Conference on Advanced Cloud and Big Data (CBD), pp. 304-309. IEEE, 2016.

[12] Y. Ni, T. Zhou, X. Gao, and H. Chen, "Design and Implementation of a UPOD for Decentralized IoT Data Platform Based on the ActivityPub Protocol," in 2023 IEEE Intl Conf on Cyber Science and Technology Congress (CyberSciTech), 2023, pp. 0825-0830.

[13] H. Chen, "Using a Ubiquitous Personal Online Datastore for Aggregating and Sharing IoT Data to Smart City," In 2022 IEEE Intl Conf on Cyber Science and Technology Congress (CyberSciTech), 2022, pp. 1-4.